\documentclass[aps,pra,amsmath,amssymb,twocolumn,superscriptaddress,notitlepage,reprint]{revtex4-2}
\usepackage[utf8]{inputenc}
\usepackage[T1]{fontenc}

\usepackage{pgfplots}
\DeclareUnicodeCharacter{2212}{−}
\usepgfplotslibrary{groupplots,dateplot,fillbetween}
\usetikzlibrary{patterns,shapes.arrows}
\pgfplotsset{compat=newest}
\tikzset{
    semithick/.style={line width=0.8pt},
}

\usepackage{scalerel,stackengine} 
\usepackage{graphicx}
\usepackage{dsfont}
\usepackage{color}
\usepackage{epsfig}
\usepackage{bm}
\usepackage{times}
\usepackage{bbold}
\usepackage{bbm}
\usepackage{subfigure}
\usepackage[colorlinks,citecolor=blue,linkcolor=blue,urlcolor=blue]{hyperref}
\usepackage{scalerel,stackengine}

\newcommand{\m}{\mathrm{m}}
\renewcommand{\c}{\mathrm{c}}
\renewcommand{\a}{\mathrm{a}}
\newcommand{\inp}{\mathrm{in}}
\newcommand{\om}{\mathrm{om}}
\newcommand{\out}{\mathrm{out}}
\newcommand{\BS}{\mathrm{\scriptscriptstyle{BS}}}
\newcommand{\DC}{\mathrm{\scriptscriptstyle{DC}}}
\newcommand{\SQL}{\mathrm{\scriptscriptstyle{SQL}}}
\newcommand{\CQNC}{\mathrm{\scriptscriptstyle{CQNC}}}
\newcommand{\ZPF}{\mathrm{\scriptscriptstyle{ZPF}}}
\newcommand{\negm}{\mathrm{\scriptscriptstyle{NMO}}}
\newcommand{\posm}{\mathrm{\scriptscriptstyle{OMS}}}
\newcommand{\Li}{\mathrm{\scriptscriptstyle{L}}}
\newcommand{\therm}{\mathrm{\scriptscriptstyle{th}}}
\newcommand{\add}{\mathrm{\scriptscriptstyle{add}}}
\newcommand{\sig}{\mathrm{\scriptscriptstyle{sig}}}
\newcommand{\esc}{\mathrm{\scriptscriptstyle{esc}}}
\newcommand{\prop}{\mathrm{\scriptscriptstyle{prop}}}
\newcommand{\sys}{\mathrm{\scriptscriptstyle{sys}}}
\newcommand{\vac}{\mathrm{\scriptscriptstyle{vac}}}
\newcommand{\bath}{\mathrm{\scriptscriptstyle{bath}}}
\newcommand{\topA}{\mathrm{NMO}\mapsto\mathrm{OMS}}
\newcommand{\topB}{\mathrm{OMS}\mapsto\mathrm{NMO}}
\newcommand{\OMS}{\mathrm{\scriptscriptstyle{OMS}}}
\newcommand{\NMO}{\mathrm{\scriptscriptstyle{NMO}}}
\newcommand{\dg}{\delta}
\newcommand{\dk}{\epsilon}
\newcommand{\xvec}{\mathbf{x}}
\newcommand{\susc}[1]{\lvert \chi_{#1} \rvert^2}
\newcommand\scalemath[2]{\scalebox{#1}{\mbox{\ensuremath{\displaystyle #2}}}}

\begin{document}

\title{All-optical coherent quantum-noise cancellation in cascaded optomechanical systems}

\author{Jakob Schweer}
\affiliation{Institute for Gravitational Physics, and Max Planck Institute for Gravitational Physics (Albert Einstein Institute), Leibniz Universit\"{a}t Hannover, Callinstra\ss{}e
  38, 30167 Hannover, Germany}

\author{Daniel Steinmeyer}
\affiliation{Institute for Gravitational Physics, and Max Planck Institute for Gravitational Physics (Albert Einstein Institute), Leibniz Universit\"{a}t Hannover, Callinstra\ss{}e
  38, 30167 Hannover, Germany}
  
\author{Klemens Hammerer}
\affiliation{Institute for Theoretical Physics, Leibniz Universit\"{a}t Hannover, Appelstra\ss{}e
  2, 30167 Hannover, Germany}
\affiliation{Institute for Gravitational Physics, and Max Planck Institute for Gravitational Physics (Albert Einstein Institute), Leibniz Universit\"{a}t Hannover, Callinstra\ss{}e
  38, 30167 Hannover, Germany}

\author{Mich\`{e}le Heurs}
\affiliation{Institute for Gravitational Physics, and Max Planck Institute for Gravitational Physics (Albert Einstein Institute), Leibniz Universit\"{a}t Hannover, Callinstra\ss{}e
  38, 30167 Hannover, Germany}
\date{\today}

\begin{abstract}
Coherent quantum noise cancellation (CQNC) can be used in optomechanical sensors to surpass the standard quantum limit (SQL). In this paper, we investigate an optomechanical force sensor that uses the CQNC strategy by cascading the optomechanical system with an all-optical effective negative mass oscillator. Specifically, we analyze matching conditions, losses and compare the two possible arrangements in which either the optomechanical or the negative mass system couples first to light. While both of these orderings yield a sub-SQL performance, we find that placing the effective negative mass oscillator before the optomechanical sensor will always be advantageous for realistic parameters. The modular design of the cascaded scheme allows for better control of the sub-systems by avoiding undesirable coupling between system components, while maintaining similar performance to the integrated configuration proposed earlier. We conclude our work with a case study of a micro-optomechanical implementation.
\end{abstract}

\maketitle

\section{Introduction}
Achieving force measurements at the quantum limit has been a significant focus for several decades \cite{braginsky1980quantum,braginsky1995quantum} and has fueled the development of optomechanics \cite{Meystre2013,Chen2013,Aspelmeyer2014}. Optomechanical sensors exploit the interaction of a light field with the motion of a mechanical oscillator to measure its displacement with high precision. Force measurements based on these schemes are subject to shot noise and quantum radiation pressure backaction noise \cite{Caves1980,Clerk2010}. Shot noise is caused by the uncertainty in the number of photons over time and can be decreased relative to the signal by increasing the intensity of the optical field. Contrary, the backaction noise arises from the fluctuation in the radiation pressure of the optical field, which will increase with its intensity. The trade-off between these competing processes then sets a lower bound to the precision of the measurement, which is called the standard quantum limit (SQL) \cite{Clerk2010,Bowen2020,Danilishin2019}.

The SQL is not a fundamental limit, and many different approaches have been suggested to achieve measurements with sub-SQL accuracy. These approaches include frequency-dependent squeezing \cite{Unruh1983,Bondurant1984,Jaekel1990}, variational measurements \cite{Vyatchanin1995,Kimble2002,Khalili2010}, dual mechanical resonators \cite{Woolley2013,Briant2003,Caniard2007,Mercier2021} and optical spring effects \cite{Buonanno2001,Verlot2010,Chen2011}. In essence, these ideas go beyond the SQL by measuring a quantum non-demolition (QND) variable of the probe, that is, a variable that commutes with itself for different moments in time. In a QND measurement, the backaction is transmitted to the canonically conjugate observable and thus avoided. A more general approach to QND measurements is gained by introducing another system that acts like a reference frame with an effective negative mass \cite{Polzik2015}. By measuring with respect to this reference system, a QND measurement is realized. When the reference system is a harmonic oscillator, an effective negative mass amounts to a negative eigenfrequency. This idea was first experimentally utilized in demonstrating Einstein-Podolsky-Rosen (EPR) states of two atomic spin oscillators of positive and negative mass \cite{Julsgaard2001}. Based on this, back action cancellation was demonstrated by Wasilewski et al. \cite{Wasilewski2010} in the context of magnetometry. Extending this idea, several proposals have been made in a hybrid setting of a mechanical oscillator and atomic spin ensembles \cite{Hammerer2009,Polzik2015}, and the evasion of backaction noise in these spin ensembles experimentally verified in \cite{Moller2017}.
Independently, Tsang and Caves \cite{Tsang2010,Tsang2012} developed this idea in a more general context, called quantum-mechanics-free subsystems. In the context of optomechanics, the main idea is to introduce an \lq anti-noise\rq{} path to the dynamics of the optomechanical sensor upon coupling to an ancillary resonator that acts as an effective negative mass. This way, the backaction noise can be cancelled coherently, and sub-SQL force sensing is achieved for all measurement frequencies. Appropriately, this approach is called coherent quantum noise cancellation (CQNC). Details and experimental feasibility of this all-optical effective negative mass oscillator were discussed in more detail by Wimmer et al. \cite{Wimmer2014}.

Within the area of CQNC force sensing, many other possible negative mass oscillators and setups have been considered. Other setups include the use of ultra-cold atoms inside a separate cavity \cite{Bariani2015,Gebremariam2019}, hybrid optomechanical cavity, i.e. implementing an atomic ensemble inside the optomechanical sensor \cite{Motazedifard2016,Singh2022} and employing Bose-Einstein condensates \cite{Zhang2013}. Even a new all-optical setup was suggested using two detuned optical modes inside the force sensor \cite{Yan2021}. These approaches can be categorized into integrated setups, where the effective negative mass is introduced directly into the optomechanical force sensor, and cascaded setups \cite{Carmichael1993,Gardiner1993}, where the effective negative mass oscillator is a separate system. Recently, Zeuthen et al. \cite{Zeuthen2022} considered a broad class of effective negative mass oscillators in a cascaded setting and even considered a possible coupling between the positive and negative mass oscillator in a parallel topology.

Inspired by this, we want to discuss a cascaded version of the original all-optical setup \cite{Tsang2010,Wimmer2014}. Instead of implementing the \lq anti-noise\rq{} path directly into the optomechanical sensor, an all-optical effective negative mass oscillator is built as a separate system. The backaction is then cancelled by coupling the force sensor to the effective negative mass oscillator via a strong coherent field. This approach will give more freedom in the experimental design and simplify reaching the challenging conditions for a CQNC experiment. The main challenge before cancelling quantum backaction noise is to measure the backaction noise. Due to the modular nature of the cascaded approach, this can be tackled entirely separate from the effective negative mass oscillator. We will see that under some modifications to the matching conditions, our cascaded setup recovers the ideal CQNC performance described in \cite{Wimmer2014}, and the additional degrees of freedom by expanding the dimension of the system lead to novel phenomena for sub-SQL force sensing. This includes the recovery of ideal CQNC around an off-resonant frequency and possible CQNC performance in the low- or high-frequency regime, even for unmatched CQNC conditions.

This paper is organized as follows. In Sec. \ref{model}, we describe the model of our cascaded CQNC scheme and derive the quantum Langevin equations of motion. In Sec. \ref{force sense}, we discuss force sensing in optomechanical sensors and derive the optimal parameters for ideal CQNC. In Sec. \ref{imp CQNC}, we analyze possible deviations from the ideal conditions and their impact on the performance of coherent quantum noise reduction. Then, in Sec. \ref{case study}, a case study is provided. Finally, we summarize our findings in Sec. \ref{conclusion}.

\begin{figure}
    \centering
    \subfigure[]{\includegraphics[width=8.6 cm]{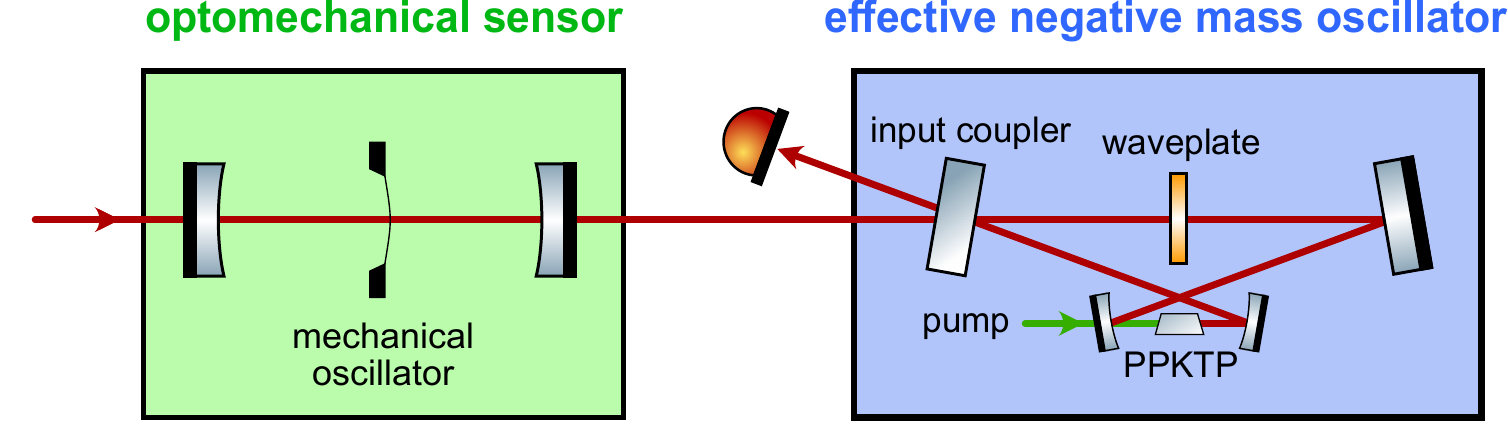}}\\
    \subfigure[]{\includegraphics[width=0.45\columnwidth]{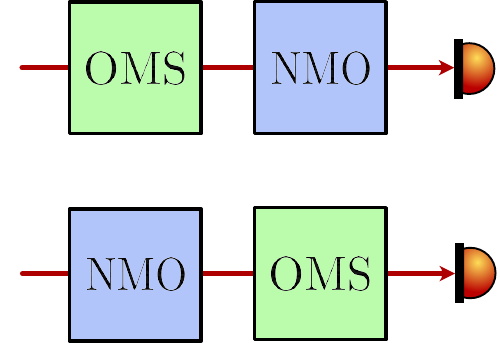}}\\
    \caption{(a)~Sketch of the cascaded setup. The optomechanical sensor (green) consists of a mechanical oscillator inside an optical cavity, depicted in a membrane-in-the-middle (MiM) setup. The effective negative mass oscillator (blue) comprises two optical modes, coupled via a beam-splitter and down-conversion process. A non-linear (PPKTP) crystal exemplifies the coupling for the down-conversion and a wave plate for the beam-splitter coupling. (b)~Simplified depiction of the possible arrangements of the cascaded scheme.}\label{fig:setup}
\end{figure}

\section{Model}\label{model}

Fig.~\ref{fig:setup}(a) illustrates a possible, schematic realization of our setup. We refer to \cite{Steinmeyer2019} for details on the experimental implementation of all-optical CQNC. An optomechanical sensor (OMS), subject to an external force and radiation pressure noise, is connected to an effective negative mass oscillator (NMO) by a coherent light field. The force is then measured by detecting outgoing light after the second system. The order of sub-systems can be chosen freely, and the two possible arrangements are depicted in Fig.~\ref{fig:setup}(b). In the first case, the light travels through the NMO, followed by the OMS. We will refer to this case as $\topA$. In the second case, the order is reversed; hence the light will travel through the OMS first, and we will refer to this case as $\topB$. The effects of these different arrangements on the performance of the backaction cancellation will be discussed later.

The OMS is modelled by an optical cavity with resonance frequency $\omega_\om$, containing a damped mechanical positive mass oscillator (PMO) with resonance frequency $\omega_\m$ and linewidth $\gamma_\m$, coupled to the cavity field via radiation pressure interaction and subjected to an external force $\mathrm{F}$. Following the standard treatment of these force sensors \cite{Aspelmeyer2014}, we move to a rotating frame with respect to the frequency $\omega_\Li$ of the strong driving laser field and arrive at the linearized Hamiltonian 
\begin{align}\label{Hampm}
\begin{split}
  H_\posm = \Delta_{\om}c^{\dagger}_{\om} c_{\om} + \frac{\omega_\m }{2}(x_\m^2+p_\m^2) + \frac{g x_\m}{\sqrt{2}} (c_\om+c^\dagger_\om).
\end{split}
\end{align}
Here, $\Delta_\om = \omega_\om-\omega_\Li$ is the detuning of the optomechanical cavity to the incoming field, $c_\om$ ($c_\om^\dagger$) are the annihilation (creation) operators of the optical mode and $x_\m = X/x_\ZPF$, $p_\m=P x_\ZPF/\hbar$ are the position and momentum operators of the mechanical oscillator normalized to the zero-point fluctuation $x_\ZPF= \sqrt{\hbar/m \omega_\m}$, such that $[x_\m,p_\m]=i$. The last term in Eq.~\eqref{Hampm} describes the radiation pressure interaction of the cavity mode and the mechanical oscillator. Its strength is given by $g= \sqrt{2}\omega_\c x_\ZPF \alpha_\c /L$, where $L$ is the cavity length and $\alpha_\c \propto \sqrt{P}$ is the field amplitude of the cavity mode, proportional to the input power $P$. Introducing dimensionless amplitude and phase quadratures $c_\om = (x_\om + i p_\om)/\sqrt{2}$, the Hamiltonian \eqref{Hampm} implies the quantum Langevin equations (QLEs)
\begin{subequations}\label{QLEpm}
\begin{align}
    \dot{x}_\m &= \omega_\m p_\m, \\
    \dot{p}_\m &= - \omega_\m x_\m - \gamma_\m p_\m - g x_\om + \sqrt{\gamma_\m} F, \\
    \dot{x}_\om &= -\frac{\kappa_\om}{2} x_\om +\Delta_{\om} p_\om + \sqrt{\kappa_\om} x_{\om}^{\inp}, \\
    \dot{p}_\om &= -\frac{\kappa_\om}{2} p_\om -\Delta_{\om} x_\om -g x_\m + \sqrt{\kappa_\om} p_{\om}^{\inp}.
\end{align}
\end{subequations}
Here, $\kappa_\om$ is the decay rate of the cavity mode and $c_\om^\inp = (x_\om^\inp + i p_\om^\inp)/\sqrt{2}$ is its vacuum input noise. The noise process fulfills $\langle c_\om^\inp (t)  c_\om^{\inp\dagger}(\tau)\rangle = \delta(t-\tau)$. In Eq.~\eqref{QLEpm} we have defined the scaled force operator $F=\mathrm{F}/\sqrt{\hbar m \gamma_\m \omega_\m}$ with dimension $\sqrt{\mathrm{Hz}}$. It consists of the to-be-detected force signal $F_\sig$ acting on the mechanical oscillator and Brownian thermal noise $F_\therm$ of the oscillator. The scaled thermal noise satisfies $\langle F_\therm(t)F_\therm(\tau)\rangle = n_\therm \delta(t-\tau)$, where $n_\therm = k_B T/\hbar \omega_\m$ is the average phonon number of the mechanical oscillator.

The NMO consists of two optical modes $c_\c$ and $a$, with resonance frequencies $\omega_\c$ and $\omega_\a$, coupled with a beam-splitter and down-conversion process. In analogy to \cite{Wimmer2014,Steinmeyer2019}, we refer to $\omega_\c$ as the meter cavity and $\omega_\a$ as the ancilla cavity. The Hamiltonian of this system, see Appendix \ref{appendixHnegm} for details, is given by
\begin{equation}\label{Hamnm}
  H_\negm = \Delta_\c c^\dagger_{\c} c_{\c} + \Delta_\a a^\dagger a +g_{\BS} (a c^{\dagger}_\c + a^\dagger c_\c) +  g_{\DC} (a c_\c+a^\dagger c^{\dagger}_\c), 
\end{equation}
with the detunings $\Delta_{\c,\a}=\omega_{\c,\a}-\omega_\Li$, the beam-splitter coupling strength $g_\BS$ and the coupling strength of the down-conversion process $g_\DC$. As above, we introduce amplitude and phase quadratures associated with the meter and ancilla cavity. Then the Hamiltonian \eqref{Hamnm} implies the following QLEs for the NMO,
\begin{subequations}\label{QLEnm}
\begin{align}
    \dot{x}_\c &= -\frac{\kappa_\c}{2} x_\c +\Delta_{\c} p_\c +(g_\BS -g_\DC) p_\a + \sqrt{\kappa_\c} x_{\c}^{\inp}, \\
    \dot{p}_\c &= -\frac{\kappa_\c}{2} x_\c -\Delta_{\c} x_\c -(g_\BS +g_\DC) x_\a + \sqrt{\kappa_\c} p_{\c}^{\inp}, \\
    \dot{x}_\a &= -\frac{\kappa_\a}{2} x_\a +\Delta_{\a} p_\a +(g_\BS -g_\DC) p_\c + \sqrt{\kappa_\a} x_{\a}^{\inp}, \\
    \dot{p}_\a &= -\frac{\kappa_\a}{2} p_\a -\Delta_{\a} x_\a -(g_\BS +g_\DC) x_\c + \sqrt{\kappa_\a} p_{\a}^{\inp}, 
\end{align}
\end{subequations}
with the cavity linewidth $\kappa_\c$ and $\kappa_\a$  and  the input noise processes $\a^\inp$ and $c_\c^\inp$.
Under the condition $g_\BS - g_\DC =0$, the QLEs \eqref{QLEnm} generate similar interaction as in Eqs.~\eqref{QLEpm}. Additionally, driving the meter cavity on resonance $\Delta_\c=0$, results in $\Delta_\a=\omega_\a - \omega_\c$. Thus the detuning between the meter and ancilla cavity can be used to generate an effective negative mass oscillator.

\section{Force sensing and ideal CQNC}\label{force sense}
To solve the dynamics of both systems, we turn to the frequency domain. Introducing the Fourier domain operators
\begin{equation*}
    O(\omega)= \frac{1}{\sqrt{2\pi}}\int \mathrm{d}t\, O(t) e^{i \omega t},
\end{equation*}
equations \eqref{QLEpm}--\eqref{QLEnm} can be solved using the standard input-output formalism \cite{Walls2007}
\begin{align}
x^\out= \sqrt{\kappa} x - x^\inp,\\
p^\out=\sqrt{\kappa} p - p^\inp.
\end{align}
We consider the systems separately, first the OMS. On resonance, $\Delta_\om=0$, the output quadratures read 
\begin{align}\label{outOM}
\begin{split}
        x^\out_\om &= e^{i \phi} x^\inp_\om,\\
        p^\out_\om &= e^{i \phi} p^\inp_\om - \chi_\m g^2 \kappa_\om \chi_\om^2 x^\inp_\om\\
        &\quad +\chi_\m \sqrt{\kappa_\om} g \chi_\om \sqrt{\gamma_\m} F,
\end{split}
\end{align}
where $e^{i \phi}=(\frac{\kappa_{\om}}{2}- i \omega )/(\frac{\kappa_{\om}}{2}+i \omega )$. We have defined the susceptibilities for the optomechanical cavity and the mechanical oscillator, as
\begin{align}
    \chi_{\om}(\omega) &= [i \omega + \frac{\kappa_\om}{2}]^{-1},\\
    \chi_\m(\omega) &= \omega_\m [(\omega^2- \omega^2_\m) - i \gamma_\m \omega]^{-1}.\label{mechsusc}
\end{align}
The mechanical oscillator is susceptible to the external force $F$, which contains the force signal $F_\sig$ and thermal noise $F_\therm$. By measuring with light, the force signal can be estimated via a phase measurement, which also introduces additional noise due to the radiation pressure. From the measured phase $p^\out_\om$ in Eq.~\eqref{outOM} we can give an unbiased estimator $\hat{F}$ of the force $F$, as 
\begin{align}\label{Fadd}
    \hat{F}= \frac{1}{\sqrt{\gamma_\m}\chi_\m\, g \sqrt{\kappa_\om}\chi_\om } p_\om^{\out}= F+F_{\add},
\end{align}
where the additional force noise, added by the measurement light, is defined as
\begin{align}
\begin{split}
    F_{\add} &= F_\therm + \frac{e^{i \phi}}{\sqrt{\gamma_\m}\chi_\m\, g \sqrt{\kappa_\om}\chi_\om } p_\om^{\inp}\\ &\quad -\frac{g \sqrt{\kappa_\om}\chi_\om }{\sqrt{\gamma_\m}} x_\om^{\inp}.
\end{split}
\end{align}
To characterize the sensitivity of the force measurement, we use the (power) spectral density of the added noise defined by
\begin{equation}
    S_{F}(\omega) \delta(\omega- \omega')= \frac{1}{2}\left(\Bigl\langle F_\add(\omega) F_\add(-\omega') \Bigr \rangle + \c.\c.\right).
\end{equation}
Assuming uncorrelated amplitude and phase quadratures, the added noise spectral density is 
\begin{equation}\label{addSQL}
    S_F = \frac{k_B T}{\hbar \omega_\m} + \frac{1}{2 G_\om \gamma_\m \lvert \chi_\m \rvert ^2} + \frac{G_\om}{2 \gamma_\m},
\end{equation}
where we defined the frequency dependent measurement strength as
\begin{align}\label{Gom}
\begin{split}
    G_\om(\omega)&= g^2 \kappa_\om \lvert \chi_\om(\omega) \rvert^2
    % =\frac{4\,g^2}{\kappa_\om}\frac{(\frac{\kappa_\om}{2})^2}{\omega^2 +(\frac{\kappa_\om}{2})^2}\\
= \Gamma_\om \;\frac{(\frac{\kappa_\om}{2})^2}{\omega^2 +(\frac{\kappa_\om}{2})^2},
\end{split}
\end{align}
with Lorentzian shape and maximum $\Gamma_\om=\frac{4\,g^2}{\kappa_\om}$. In this form, the noise spectral density \eqref{addSQL} is dimensionless. To arrive at a force noise spectral density in units of $\mathrm{N}^2/\mathrm{Hz}$, one has to re-scale it, such that $S_F(\omega)= \hbar m \gamma_\m \omega_\m S_F (\omega)$ for a given optomechanical force sensor \cite{Wimmer2014}. 
The terms in Eq.~\eqref{addSQL} are thermal noise due to Brownian motion of the mechanical oscillator (first term), shot noise in the phase quadrature (second term) and backaction noise from the amplitude quadrature (third term). The thermal noise adds a flat background to the force sensitivity, which is independent of the measurement rate $G_\om$ and frequency.
Throughout this paper, we will assume that the thermal noise is either dominated by backaction noise \cite{Wimmer2014} or is suppressed by cooling of the mechanical mode and therefore neglect this term. 

The shot noise term scales inversely proportional to the measurement rate $G_\om$ and thus inversely proportional to the power, as $G_\om \propto P$. Additionally, the backaction noise is proportional to $G_\om$. This implies that an optimal power value, which minimizes Eq.~\eqref{addSQL}, exists for each frequency. Minimizing Eq.~\eqref{addSQL} with respect to $G_\om$ for all frequencies gives the achievable lower bound 
\begin{equation}\label{SQL}
    S_F(\omega) \geq \frac{1}{\gamma_m \lvert \chi_\m(\omega) \rvert} \equiv S_\SQL(\omega),
\end{equation}
which is the standard quantum limit (SQL). The optimal measurement rate required is 
\begin{equation}
    G_\SQL(\omega) = \frac{1}{\lvert \chi_\m(\omega)\rvert}.
\end{equation}

Next, we consider the NMO. On resonance, $\Delta_\c=0$, and for $g_\BS=g_\DC=\frac{1}{2} g_\a$ the output quadratures read   
\begin{align}\label{outNM}
\begin{split}
        x^\out_\c &= e^{i \theta} x^\inp_\c,\\
        p^\out_\c &= e^{i \theta} p^\inp_\c - \chi_\a g_\a^2 \kappa_\c \chi_\c^2 x^\inp_\c\\
        &\quad +\chi_\a \sqrt{\kappa_\c} g_\a \chi_\c \sqrt{\kappa_\a} \left(\frac{\kappa_\a/2+i\omega}{\Delta_\a} x_\a^\inp + p_\a^\inp\right),
\end{split}
\end{align}
where $e^{i \theta}=(\frac{\kappa_{\c}}{2}- i \omega )/(\frac{\kappa_{\c}}{2}+i \omega )$. We have defined susceptibilities of the meter cavity $\chi_\c$ and the ancilla cavity $\chi_\a$ as
\begin{align}
    \chi_{\c} &= [i \omega + \frac{\kappa_\c}{2}]^{-1}, \\
    \chi_\a &= \Delta_\a [(\omega^2-\Delta_\a^2-\frac{\kappa_\a^2}{4})-i \kappa_\a \omega]^{-1}\label{ancsusc}.
\end{align}
The aim of our setup is to couple the two systems in a manner that the backaction noise in the force spectrum will cancel, hence allowing for a sub-SQL performance. In our dual-cavity setup, this is done by cascading the two systems and matching the parameters of the NMO accordingly. 

As seen in Fig.~\ref{fig:setup}(b), the whole scheme has two possible arrangements. Thus, to cascade the systems, we choose  $x_\c^\out=x_\om^\inp$ and $p_\c^\out=p_\om^\inp$ for the case $\topA$ and  $x_\om^\out=x_\c^\inp$ and $p_\om^\out=p_\c^\inp$ for $\topB$. For ideal CQNC, the order will not matter. We will discuss cases that depend on the order further below. After cascading the two systems, we can again identify the additional force noise as in Eq.~\eqref{Fadd} and derive the added noise spectral density
\begin{align}\label{cascSf}
    \begin{split}
      S_F  &= \frac{1}{2 G_{\om} \gamma_\m |\chi_\m|^2 }\\ 
    % &+\frac{|g_\a^2 \kappa_\c \chi_\c^2 \,\chi_\a\, e^{i \phi} + g^2  \kappa_{\om} \chi_{\om}^2\, \chi_\m \,e^{i\theta}|^2}{2 G_{\om} \gamma_\m |\chi_\m|^2 }\\
    &\quad+\frac{G_\a^2 \susc{\a}+ G_{\om}^2 \susc{\m} +2\,G_\a G_\om \Re(\chi_\m \chi_\a^*)}{2 G_{\om} \gamma_\m |\chi_\m|^2 }\\
    &\quad+ \frac{G_{\a} \kappa_\a |\chi_{\a}|^2 }{2 G_{\om} \gamma_\m |\chi_\m|^2}\left(\frac{\omega^2 + \kappa_\a^2/4}{\Delta_\a^2}+1\right),
    \end{split}
\end{align}
see Appendix \ref{appendixnoiseden} for details.
Analogous to the measurement strength of the OMS in Eq.~\eqref{Gom}, we have defined the frequency dependent measurement strength of the NMO,
\begin{align}\label{Ga}
\begin{split}
    G_\a(\omega) &=g_\a^2 \kappa_\c \lvert \chi_\c(\omega) \rvert^2
    % = \frac{4\,g_\a^2}{\kappa_\c}\frac{(\frac{\kappa_\c}{2})^2}{\omega^2 +(\frac{\kappa_\c}{2})^2}\\
    = \Gamma_\a \;\frac{(\frac{\kappa_\c}{2})^2}{\omega^2 +(\frac{\kappa_\c}{2})^2},
\end{split}
\end{align}
with $\Gamma_\a=\frac{4\,g_\a^2}{\kappa_\c}$.
The terms in Eq.~\eqref{cascSf} are shot-noise (first term), backaction noise (second term) and shot noise from the ancilla cavity (last term). The backaction noise is then cancelled if the conditions are such that

\begin{align}
    g_\BS &= g_\DC = \frac{1}{2} g_\a\label{DCBSmatch},\\
    G_\om(\omega) &= G_\a(\omega)\label{matchmeasure},\\
   \chi_\m(\omega) &= - \chi_\a(\omega)\label{matchingsusc},
\end{align}
for all $\omega$. This means that the ancilla cavity should couple to the light with the same strength as the PMO, but the response to the force signal should be opposite to the PMO, hence it behaves as an effective negative mass. Considering the explicit form of Eqs.~\eqref{mechsusc} and \eqref{ancsusc}, condition \eqref{matchingsusc} entails further restrictions:
\addtocounter{equation}{-1}
\begin{subequations}
\begin{enumerate}
    \item The detuning of the ancilla cavity to the meter cavity is
    \begin{equation}
        \Delta_\a = - \omega_\m,
    \end{equation}
    which effectively moves the ancilla cavity to the negative mass frame.
    \item The linewidth of the ancilla cavity $\kappa_\a$ should match the damping rate of the mechanical oscillator
    \begin{equation}
        \kappa_\a = \gamma_\m,
    \end{equation}
    to mimic the oscillating behaviour of the PMO.
    \item The susceptibilities $\chi_\m$ and $\chi_\a$ differ by a factor $\kappa_\a^2/4$. To alleviate this, the detuning $\lvert\Delta_\a\rvert \gg \kappa_\a$, and together with forgoing points this implies the resolved sideband limit of the ancilla cavity
    \begin{equation}
        \omega_\m \gg \kappa_\a,
    \end{equation}
    and a large quality factor of the mechanical oscillator,
    \begin{equation}
        Q_\m=\frac{\omega_\m}{\gamma_\m} \gg 1.
    \end{equation}    
\end{enumerate}
\end{subequations}
These conditions are similar to the integrated setup \cite{Wimmer2014}, but instead of the coupling strength $g_\a$ and $g$, the measurement strengths $G_\a$ and $G_\om$ need to match. 

Assuming conditions \eqref{DCBSmatch}--\eqref{matchingsusc} are met, the backaction term in Eq.~\eqref{cascSf} will vanish, and we arrive at
\begin{equation}
    S_F = \frac{1}{2 G_{\om} \gamma_\m |\chi_\m|^2 } + \frac{1}{2}\left(\frac{\omega^2 + \gamma_\m^2/4}{\omega_\m^2}+1\right),
\end{equation}
which contains only shot-noise contributions of the measured phase quadrature and the ancilla cavity. The contribution of the OMS is sometimes referred to as the fundamental quantum limit (FQL)\cite{Danilishin2019}, energetic quantum limit \cite{Braginsky2003} or the quantum Cramér-Rao bound \cite{Tsang2011, Miao2017,Khalili2021}. In the limit of large measurement strength, we arrive at the lower bound
\begin{align}\label{idealCQNC}
    S_F(\omega) \geq \frac{1}{2}\left(\frac{\omega^2 + \gamma_\m^2/4}{\omega_\m^2}+1\right) \equiv S_\CQNC(\omega).
\end{align}
Combining Eqs.~\eqref{SQL} and \eqref{idealCQNC} we find
\begin{align}
    S_\CQNC = S_\SQL\times\frac{1}{2 Q_\m}\left(\frac{\omega^2 + \gamma_\m^2/4+\omega_\m^2}{\sqrt{(\omega^2-\omega_\m^2)^2+\gamma_\m^2\omega^2}}\right).
\end{align}
Thus, for $Q_\m \gg 1$, we can summarize
\begin{align}
    S_\CQNC = S_\SQL \times \begin{cases} \quad 1 &\text{on resonance } \omega=\omega_\m\\
  1/(2 Q_\m) & \text{off resonance } \omega\,\substack{\ll\\\gg}\,\omega_\m.
    \end{cases}
\end{align}
In conclusion, under the additional condition that $G_\om = G_\a$, the cascaded setup reproduces the same findings of \cite{Wimmer2014}, leading to an enhancement in performance up to a factor of $2 Q_\m$ off-resonance and SQL performance on resonance.

\section{Imperfect CQNC}\label{imp CQNC}
Conditions \eqref{DCBSmatch}--\eqref{matchingsusc} are the ideal case for a perfect cancellation of backaction noise and will not be satisfied in an actual experiment. Therefore, we will discuss possible imperfections and their impact on the performance of our cascaded scheme. These imperfections include mismatches to the parameters in Eqs.~\eqref{DCBSmatch}--\eqref{matchingsusc}, and possible losses. Another degree of freedom of our setup is the order in which the light passes through the sub-systems (i.e. $\topA$ or $\topB$), but this will only affect the force sensing performance for imperfections that directly affect the force signal. Hence, we split our discussion into order dependent and independent categories. Data shown in the figures of this section refer to an OMS given by the parameters in Table~\ref{OMSpar}.

\begin{table}
\caption{Parameters of the optomechanical sensor used in Figs.~\ref{fig2} to \ref{relcoupl}}
\label{OMSpar}
\begin{tabular}{l l c c}
\toprule
\multicolumn{2}{l}{Parameter}&  Norm. value &  Value\\
\hline
$\omega_\m$&mechanical resonance frequency  & 1 & 500 kHz \\
$\gamma_\m$&mechanical linewidth & $10^{-3}\,\omega_\m$ & 500 Hz \\
$Q_\m$&mechanical quality factor & $\frac{\omega_\m}{\gamma_\m}$ & 1000 \\
$\kappa_\om$&optomechanical cavity linewidth& $10\,\omega_\m$ & 5 MHz \\
\hline
\hline
\end{tabular}
\end{table}

\subsection{Order-independent imperfections}
The parameters discussed in this subsection will impede the cancellation of backaction noise and, as a result, limit the CQNC performance but will not affect the force signal. Hence, the possible CQNC performance in the face of these imperfections will not depend on the system order.

\subsubsection{Non-ideal ancilla cavity linewidth \texorpdfstring{$\kappa_\a \neq \gamma_\m$}{ka not gm}}
The strictest requirement for an all-optical CQNC setup is to match the ancilla cavity linewidth to the damping rate of the mechanical oscillator. Assuming all conditions for ideal CQNC are matched, except $\kappa_\a\neq\gamma_\m$. Since the measurement strengths, $G_\om = G_\a$ are matched for all frequencies, and we assume no propagation losses between the system, this effectively reduces to the integrated CQNC setup \cite{Wimmer2014}. The spectral density of added noise \eqref{cascSf} in this case becomes

\begin{align}\label{miskappa}
\begin{split}
S_F &=  \frac{1}{2 G_{\om} \gamma_\m |\chi_\m|^2 }
+\frac{G_\om}{2 \gamma_\m} \left\lvert \frac{\chi_\m +\chi_\a}{\chi_m}\right\rvert^2 \\
&\quad+\frac{\kappa_\a \lvert \chi_\a \rvert^2 }{2 \gamma_\m  \lvert \chi_\m \rvert^2 }\left(\frac{\omega^2 + \kappa_\a^2/4}{\omega_\m^2}+1\right).
\end{split}
\end{align}

\begin{figure}
\includegraphics{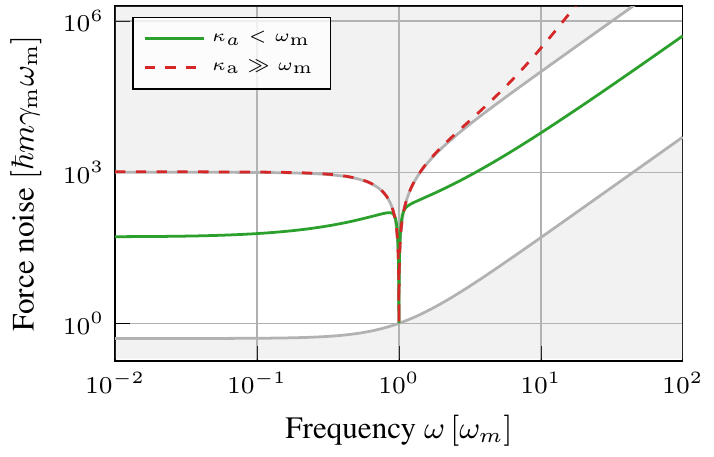}\\
  \caption{Force noise for a mismatch between ancilla cavity linewidth $\kappa_\a$ and damping rate of the mechanical oscillator $\gamma_\m$. For $\kappa_\a < \omega_\m$ an improvement of $\kappa_\a/2\omega_\m$ can be achieved off-resonance (solid green). For $\kappa_\a \gg \omega_\m$, the effect of CQNC is completely cancelled for low frequencies, and the sensitivity is worse than the SQL for high frequencies (red). The shaded areas mark the bounds for sub-SQL sensitivity, from below the fundamental limit given by Eq.~\eqref{idealCQNC}, from above the SQL given by Eq.~\eqref{SQL}. Parameters are given in Table~\ref{OMSpar}}\label{fig2}
\end{figure}

For an optimal $G_\om$, we find the minimal spectral density for the added noise,
\begin{align}\label{diffgammaSf}
    S_F = \frac{\lvert \chi_\m + \chi_\a \rvert}{\gamma_\m \lvert \chi_\m \rvert} 
    + \frac{\kappa_\a \lvert \chi_\a \rvert^2 }{2 \gamma_\m  \lvert \chi_\m \rvert^2 }\left(\frac{\omega^2 + \kappa_\a^2/4}{\omega_\m^2}+1\right)
\end{align}
This is composed of measurement shot and backaction noise (first term) and noise introduced by the ancilla cavity (second term). The second term will dominate the first one for frequencies off-resonance, setting a bound to the achievable performance. The ratio between the spectral density \eqref{diffgammaSf} and the SQL is
\begin{align}\label{miskappaLim}
    S_F = \frac{\kappa_\a}{2\,\omega_\m} \times S_\SQL,
\end{align}
for $\kappa_\a < \omega_\m$. For $\kappa_\a \gg \omega_\m$, the effect of CQNC will vanish for low frequencies, converging to the SQL, while for large frequencies, the added noise is larger than the SQL.
This is illustrated in Fig.~\ref{fig2}.

\subsubsection{Unequal measurement strengths \texorpdfstring{$G_\om \neq G_\a$}{Gom not Ga}}
Next, we consider a mismatch of the measurement strengths $G_\a \neq G_\om$, while matching the other CQNC conditions. This entails unmatched cavity linewidth $\kappa_\c\neq\kappa_\om$ and unmatched couplings $g_\a \neq g$. Introducing parameters for the linewidth mismatch $\kappa_\c = \dk\, \kappa_\om$ and coupling mismatch $g_\a = \sqrt{\dg} g$, we find for the spectral density
\begin{align}\label{diffrateSf}
\begin{split}
     S_F &= \frac{1}{2 G_{\om} \gamma_\m |\chi_\m|^2 } 
    + \frac{G_\om}{2 \gamma_\m} \left\lvert 1- \delta \dk \frac{\lvert \chi_\c \rvert^2}{\lvert \chi_\om \rvert^2}\right\rvert^2 \\
&\quad+\delta \dk \frac{\lvert \chi_\c \rvert^2}{\lvert \chi_\om \rvert^2}\frac{1}{2}\left(\frac{\omega^2 + \gamma_\m^2/4}{\omega_\m^2}+1\right).   
\end{split}
\end{align}
For suitable couplings $g$ and cavity linewidth $\kappa$, we can find a frequency where the backaction term in Eq.~\eqref{diffrateSf} will vanish, and ideal CQNC is possible. This is the case when the Lorentzians $G_\a$ and $G_\om$ are such that they will intersect at a frequency $\omega \neq 0$.
We find that
\begin{align}\label{newdip}
    \omega_{*} = \pm \sqrt{\frac{\dg\,\dk - \dk^2}{1-\dg\,\dk}}\frac{\kappa_\om}{2}
\end{align}
is a real-valued frequency for the following parameters:
\begin{subequations}
\begin{align}
g_\a = g& \quad \Rightarrow\; & \kappa_\c&< \kappa_\om &\mathrm{or}& &\kappa_\c &> \kappa_\om,\\
g_\a < g& \quad \Rightarrow\; & \dk&> \frac{1}{\dg} &\mathrm{or}& &\dk&<\dg,\\
g_\a > g& \quad \Rightarrow\; & \dk&>\dg &\mathrm{or}& &\dk&<\frac{1}{\dg}.
\end{align}
\end{subequations}
Consequently, a cavity linewidth mismatch can compensate for every possible matching condition of the couplings, and ideal CQNC can be achieved at $\omega_*$. 

For non-vanishing backaction, we can again minimize the spectral density \eqref{diffrateSf} with an optimal $G_\om$. Turning to the low-frequency limit ($\kappa_{\c,\om} \gg \omega$), the measurement strengths become frequency independent, $G_{\om,\a} \to \Gamma_{\om,\a}$, and the ratio $\lvert \chi_\c \rvert^2/\lvert \chi_\om \rvert^2 \to 1/\dk^2$.  The minimal noise spectral density is then
\begin{align}
    S_F = |1- \frac{\dg}{\dk}| \times  S_\SQL + \left(\frac{\dg}{\dk}\right) \times S_\CQNC.
\end{align}
Ideal CQNC can be recovered for $\dk=\dg$, which means $\Gamma_\om=\Gamma_\a$. Hence, as long as the rate at which the backaction information leaks out of the system is matched, ideal CQNC is possible.

\begin{figure}
\centering
\includegraphics{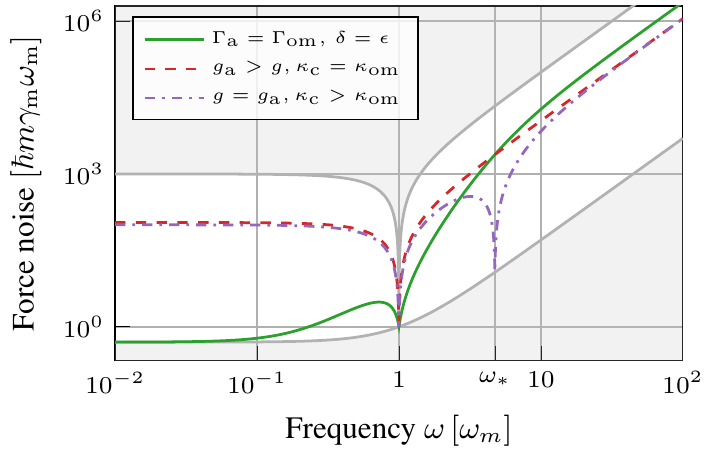}\\
  \caption{Force noise for imperfect matching of measurement strength. For mismatched coupling strength compensated by linewidth mismatch, perfect noise cancellation can be recovered at low frequencies (solid green for $\dk=\dg=0.9$). For matched linewidth but mismatched coupling strength, noise cancellation is limited, but sub-SQL performance is possible (dashed red for $\dg=0.9$). In the case of matched coupling strength but mismatched linewidth, we find a frequency \eqref{newdip} where perfect noise cancellation is possible (dash-dotted purple line, with $\dk=0.9$). The shaded areas mark the bounds for sub-SQL sensitivity, from below the fundamental limit given by Eq.~\eqref{idealCQNC} and from above the SQL given by Eq.~\eqref{SQL}. Parameters are given in Table~\ref{OMSpar}.}\label{fig3}
\end{figure}

For the converse case ($\kappa_{\c,\om} \ll \omega$) the cavity susceptibilities $\susc{\c} \approx \susc{\om}$, effectively cancelling in Eq.~\eqref{diffrateSf}. The optimal spectral density becomes 
\begin{align}
    S_F =  |1- \dg \,\dk| \times  S_\SQL + \left(\dg \,\dk\right) \times S_\CQNC.
\end{align}
In this case, ideal CQNC can be recovered for $\dg=1/\dk$, which entails $g_\a^2 \kappa_c = g^2 \kappa_\om$. We have depicted our main findings in Fig.~\ref{fig3}. 

We also considered a combination of the imperfections discussed in this section. If, for example, $G_\om \neq G_\a$ and additionally $\kappa_\a \neq \gamma_\m$, the noise spectral density will be a combination of Eqs.~\eqref{miskappa} and \ref{diffrateSf}. In this case, the cancellation of backaction noise is possible for the cases discussed above, but the ancilla cavity noise floor is higher because of the linewidth mismatch $\gamma_\m\neq \kappa_\a$. Thus, our findings will remain the same, but the achievable performance off-resonance is given by the noise spectral density \eqref{miskappaLim}.    

\subsection{Order-dependent imperfections}\label{orderdep}

The parameters discussed in this subsection, namely losses, not only hamper the cancellation of backaction noise but also affect the force signal directly.

\subsubsection{Losses}
We first consider propagation losses, which occur between the first and the second system. The propagation losses are modelled via mixing the output signal of the first system with vacuum in a beam-splitter-like interaction. This leads to a modified output signal \cite{Bowen2020},
\begin{align}
    x_\out^{\prime} = \sqrt{\eta}\,x_\out + \sqrt{1-\eta}\,x_\vac, 
\end{align}
where $x_\vac$ represents the vacuum field and $\eta \in [0,1]$ is the efficiency of the process. Due to this additional noise, information about the backaction interaction of the first system is lost to the vacuum, hence perfect cancellation of backaction noise is not possible. As before, we can find an optimal coupling strength to minimize the additional noise. For the system order $\topA$, we achieve a minimal spectral density off-resonance
\begin{align}\label{proplossA}
    S_F = \sqrt{1- \eta} \times S_\SQL.
\end{align}
In the opposite order $\topB$, additionally to the loss of backaction information, some force signal will be lost due to propagation losses. Hence, the added noise will increase for this topology. We find 
\begin{align}\label{proplossB}
       S_F= \frac{\sqrt{1-\eta}}{\eta} \times S_{\SQL},
\end{align}
for the minimal spectral density off-resonance. The spectral density is increased by $1/\eta$ compared to the case $\topA$ and hence directly proportional to the lost force signal. 
Losses after the second system constitute the detection efficiency and can be modelled similarly. Since this will not affect the cancellation of backaction noise, we will omit detection losses for now.

Apart from propagation losses, we take intracavity losses into account. Introducing a Markovian bath for each cavity, with coupling rates $\kappa_\c^\bath$ and $\kappa_\om^\bath$, the intracavity losses can be described in terms of the escape efficiencies 
\begin{align}
    \eta_{\om,\c}^\esc = \frac{\kappa^{\inp}_{\om,\c}}{\kappa^{\inp}_{\om,\c}+\kappa^{\bath}_{\om,\c}}=   \frac{\kappa^{\inp}_{\om,\c}}{\kappa_{\om,\c}}.
\end{align}
Similar to propagation losses, introducing intracavity losses will always impede the cancellation of backaction noise, and depending on the order of the systems, the available force signal information will differ. For the case $\topA$, with optimal measurement strength, we find the minimal spectral density
\begin{align}\label{intlossA}
     S_F = \frac{\sqrt{\eta^\esc_c + \eta^\esc_{\om}-2 \eta^\esc_c \eta^\esc_{\om}}}{\eta^\esc_{\om}}\times S_{\SQL}.
\end{align}
This encompasses both cases with propagation loss, for $\eta^\esc_\om \to 1$ we retrieve Eq.~\eqref{proplossA} and for $\eta^\esc_\c \to 1$ Eq.~\eqref{proplossB} respectively. Thus, for the configuration $\topA$, the intracavity loss can be handled similarly to propagation loss.

For the case $\topB$, we lose additional force signal due to intrinsic loss in the meter cavity. Moreover, the signal also picks up additional information of the phase quadrature. We arrive at the minimal spectral density,
\begin{align}\label{intlossB}
S_F &= \frac{\sqrt{\eta^\esc_\c + \eta^\esc_{\om}-2 \eta^\esc_\c \eta^\esc_{\om}}}{\eta^\esc_{\om}\,|1-\eta^\esc_\c \kappa_\c \chi_\c |^2} \times S_{\SQL}.
\end{align}
The term $|1-\eta^\esc_\c \kappa_\c \chi_\c |^2$ describes the meter cavity's phase and noise contribution. Due to its dependence on the meter cavity susceptibility $\chi_\c$, this difference is frequency-dependent and will vanish for frequencies $\omega > \kappa_c$. For low frequencies, it will be at a maximal value of $|1-2 \eta^\esc_\c|^2$, making intracavity losses extra punishing for configuration $\topB$. 
% Our expressions from Eqs. (\ref{proplossA})-(\ref{intlossB}) can be combined into a lengthy, yet simple, expression to encompass all the losses into a total loss expression. If we consider $1\%$ losses from the sources above, we arrive at an optimal spectral density of
% \begin{align}
%     S_F \approx S_\SQL \times \begin{cases} 17.4\%& \text{for }\topA \\
%  18.3\%& \text{for }\topB
% \end{cases},
% \end{align}
% for low frequencies.
We see that introducing losses is detrimental to the possible noise reduction. As losses will never be avoidable, the system order $\topA$ should always be preferable since higher levels of noise reduction are achieved.

% \begin{figure}
%      \centering
%      \begin{subfigure}
%          \centering
%          \input{TikZ/proploss}\\
%      \end{subfigure}
%      \begin{subfigure}
%          \centering
%          \input{TikZ/intloss}
%      \end{subfigure}
%      \caption{Force noise for propagation and intrinsic losses. (a) Losses between the first and second system prevent perfect noise cancellation, imposing a limit to the CQNC performance (solid green). If the light enters the OMS first, some additional force signal will be lost, further degrading the performance (dashed red). Plotted with efficiency $\eta=90\%$. (b) Intrinsic losses in the cavities limit the noise cancellation (solid green), and for the case $\topB$, additional force signal is lost due to the lossy NMO. This difference is frequency dependent and will vanish for high frequency. Plotted for escape efficiencies $\eta^\esc_\c=\eta^\esc_\om=95\%$. The shaded areas mark the bounds for sub-SQL sensitivity, from below the fundamental limit given by Eq. (\ref{idealCQNC}) and from above the SQL given by Eq. (\ref{SQL}).}\label{losses}
% \end{figure}

\subsubsection{Relative mismatch of \texorpdfstring{$g_\BS$}{gBS} and \texorpdfstring{$g_\DC$}{gDC}}

\begin{figure}
\includegraphics[width=0.4\columnwidth]{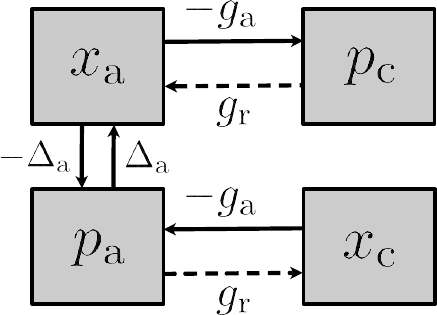}\\
  \caption{Flow chart between the mode of the ancilla and the meter cavity. The solid line shows the original backaction flow, and the dashed line shows the noise introduced by the relative mismatch $g_\mathrm{r}$ of the beam-splitter and down-conversion coupling.}\label{flow chart}
\end{figure}

\begin{figure}
     \begin{subfigure}
        \centering
        \includegraphics{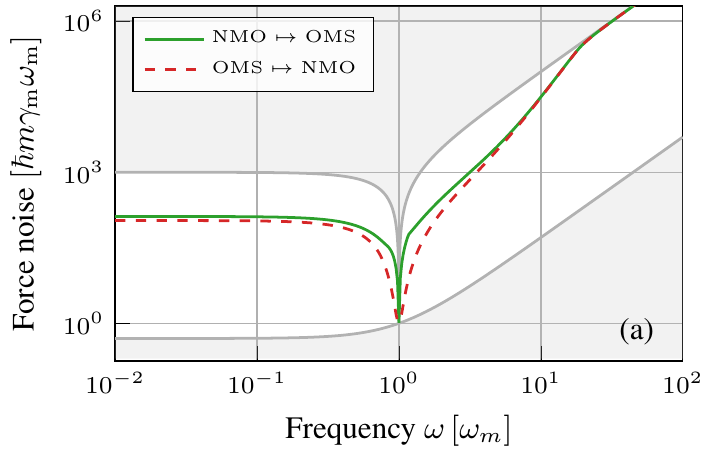}\\
     \end{subfigure}
     \begin{subfigure}
        \centering
        \includegraphics{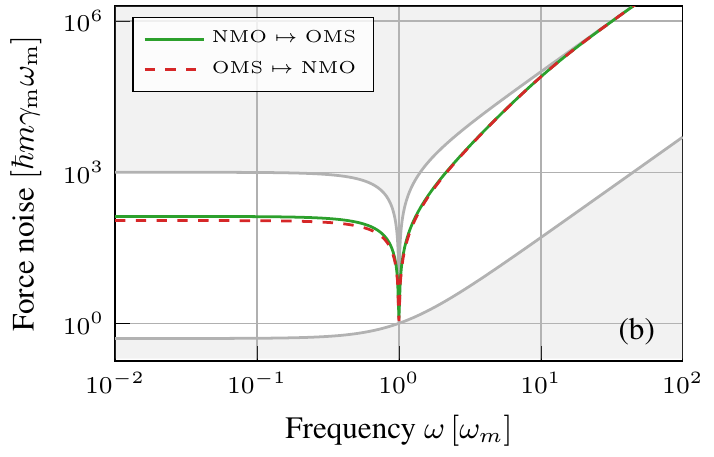}\\
     \end{subfigure}
     \caption{Force noise for relative mismatch of beam-splitter and down-conversion coupling. (a) Plotted for $g_\mathrm{r}=0.2$. (b) Plotted for $g_\mathrm{r}=-0.2$. The relative mismatch introduces additional noise by modifying the effective backaction term of the NMO. Perfect noise cancellation is not possible, but sub-SQL levels are achievable for low frequencies. For high frequency, no noise reduction is possible. Traversing through the OMS first seems advantageous for noise cancellation. The shaded areas mark the bounds for sub-SQL sensitivity, from below the fundamental limit given by Eq.~\eqref{idealCQNC} and from above the SQL given by Eq.~\eqref{SQL}. Parameters are given in Table~\ref{OMSpar}.}\label{relcoupl}
\end{figure}

In addition to losses, a relative mismatch between the beam-splitter coupling $g_\BS$ and down-conversion coupling $g_\DC$ will also affect the noise cancellation depending on the system order. So far, we assumed $g_\BS = g_\DC =1/2\, g_\a$, in order to mimic the backaction interaction of the OMS. We will now fix $g_\BS + g_\DC = g$ and introduce a mismatch between the beam-splitter and down-conversion couplings 
\begin{align}
    \frac{g_\BS-g_\DC}{g_\BS + g_\DC}= g_\mathrm{r} \neq 0.
\end{align} 
As shown in Fig.~\ref{flow chart}, the relative mismatch $g_\mathrm{r}$ allows the phase quadratures to couple back into the amplitude quadrature and thus deviate from the backaction interaction of the OMS. This introduces a noise path and will limit the cancellation of backaction noise. It also affects the force noise differently for the different system orders. For the case $\topB$, the force signal is imprinted on the output phase quadrature of the OMS, and with the introduced mismatch, it is possible for the signal to couple to the amplitude quadrature. Contrary, for $\topA$, the force signal will remain fully in the output phase quadrature. Thus, this results in different spectral noise densities for our phase measurement. For general mismatches, this will not reduce to a simple expression. The resulting spectral densities were calculated numerically and are shown in Fig.~\ref{relcoupl}. The CQNC performance is limited for low frequencies, but sub-SQL levels are still possible. Contrary to losses, the order $\topB$ seems advantageous for a relative mismatch of the couplings. CQNC will vanish entirely in the high-frequency limit, and no sub-SQL performance is possible.

\section{case study}\label{case study}

After discussing ideal CQNC and the most relevant deviations from the ideal parameters, we now turn to a realistic situation one would expect in an actual experiment. For an integrated setup, reasonable parameters have been discussed in \cite{Wimmer2014}, which were revised in \cite{Steinmeyer2019} for a cascaded setup, and two reasonable sets of parameters were given. From there, we found a new set of parameters which achieve broadband noise reduction for frequencies below the mechanical resonance of the oscillator. Losses are of particular interest in our case study, as they influence the noise reduction depending on the system order. Our set of parameters is shown in Table~\ref{parametertable}. 

\begin{table}
\caption{Proposed set of parameters}
\label{parametertable}
\begin{tabular}{l l c c}
\toprule
\multicolumn{2}{l}{Parameter}&  Norm. value &  Value/$2\pi$\\
\hline
$\omega_\m$&mechanical resonance frequency  & 1 & 500 kHz \\
$\gamma_\m$&mechanical linewidth & $10^{-8}\,\omega_\m$ & 5 mHz \\
$\kappa_\c$&meter cavity linewidth& $4\,\omega_\m$ & 2 MHz\\
$\Delta_\a$&ancilla cavity detuning  & $-0.99\,\omega_m$ & -495 kHz\\
$\kappa_\a$& ancilla cavity linewidth  & $\frac{2}{5}\,\omega_\m$ & 200 kHz\\
$g_\BS$& beam-splitter coupling strength  & $1.01\,\frac{g}{2}$ & 253 kHz\\
$g_\DC$ & down-conversion coupling strength  & $0.97\,\frac{g}{2}$ & 243 kHz\\
$\eta^\esc_\c$& escape efficiency NMO  & $0.9$ & \\
$\kappa_\om$&optomechanical cavity linewidth& $0.99\,\kappa_\c$ & 1.98 MHz \\
$g$& optomechanical coupling strength  & $\omega_m$ & 500 kHz\\
$\eta^\esc_\om$& escape efficiency OMS  & $0.9$ & \\
$\eta_\prop$& propagation efficiency& $0.97$ & \\
$\eta_{\det}$& detection efficiency & $0.97$ & \\
\hline
\hline
\end{tabular}
\end{table}

The OMS must be limited by quantum backaction noise to measure the possible cancellation of backaction noise. For this, the quantum backaction noise in Eq.~\eqref{addSQL} must be much larger than the thermal noise. In the low-frequency limit $(\kappa_\om\gg\omega)$, this can be expressed in terms of the quantum cooperativity, as
\begin{align}
    C_\mathrm{q} = \frac{\Gamma_\om}{\gamma_\m} \frac{\hbar\,\omega_\m}{k_{B} T}=\frac{4 g^2 \hbar}{\kappa_\om k_{B} T}\,Q_\m \gg 1.
\end{align} 
Modern silicon-nitride membranes have exceeded quality factors of $Q_\m=10^8$ \cite{Mason2019}, thus the OMS would be quantum backaction limited for a temperature $T=4\text{ K}$, a temperature achievable with cryogenics. For higher temperatures, the quality factor must be increased to elevate the backaction effects over the thermal noise floor, and similarly, lower temperatures allow for a lower quality factor. In order to account for this and compare all OMS of frequency $\omega_\m$, once they can resolve the quantum backaction, we normalize our force noise by the quality factor $Q_\m$.

Matching most parameters, such as ancilla cavity detuning $\Delta_\a$ and the cavity linewidth $\kappa_\c$ and $\kappa_\om$ should not be a problem; we assume them to be closely matched. More delicate to match are the coupling strengths. A down-conversion coupling of $g_\DC=2\pi\times 250\text{ kHz}$ and a beam-splitter coupling of $g_\BS \geq2 \pi \times 235\text{ kHz}$ were readily achieved \cite{Steinmeyer2019}, thus we set the optomechanical coupling strength to $g=2 \pi \times  500\text{ kHz}$. Optomechanical coupling strengths of $g=2 \pi\times 440 \text{ kHz}$ have been reported in micro-mechanical setups \cite{Moller2017} and higher couplings in the order of MHz should be possible \cite{Norte2016}. Hence, our assumed coupling strength should be reasonable. If these levels cannot be reached for the optomechanical coupling strength, one could still compensate for this mismatch by the cavity linewidths, as described in Eq.~\eqref{diffgammaSf} and increase the performance for low frequencies.

For a negative mass oscillator, where the two modes are not spatially separated, as depicted in Fig.~\ref{fig:setup}(a), the escape efficiency will also dictate the achievable linewidth of the ancilla cavity. An escape efficiency of 90\% should be achievable \cite{privateJonas}, which, with a meter cavity linewidth of $\kappa_\c= 2\text{ MHz}$, makes an ancilla cavity linewidth of $\kappa_\a = 200 \text{ kHz}$ possible. For the OMS, similar escape efficiencies should be achievable. Detection efficiencies over $97\%$ were already realized \cite{Vahlbruch2016}. Similarly, propagation losses between the systems should not be an issue. We assume $3\%$ losses from both propagation and detection.

\begin{figure}
    \includegraphics{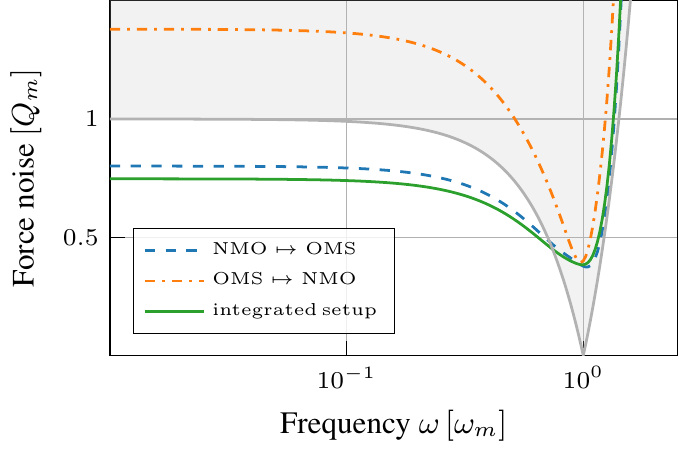}\\
    \caption{Force noise normalised to $Q_\m$ for the parameters given by Table~\ref{parametertable} and temperature $T=4\text{ K}$. For low frequencies, sub-SQL performance is possible for the integrated setup (solid green) and the case $\topA$ (dashed blue). No sub-SQL levels are possible for the case $\topB$ (dash-dotted orange). The shaded area shows levels above the SQL.}\label{realistic}
\end{figure}

The achievable sensitivities for the parameters in Table~\ref{parametertable} are shown in Fig.~\ref{realistic}. In the low-frequency regime, the configuration $\topA$ shows a reduction of 20\% below the SQL and almost comparable results to the integrated setup. No sub-SQL sensitivity can be achieved for the other system order $\topB$. This is not surprising, as we saw in subsection \ref{orderdep} that this configuration suffers additional penalties from losses. We see that instead of matching the parameters \eqref{DCBSmatch}--\eqref{matchingsusc},  the limiting factor for noise reduction in a realistic case will be losses. Additionally, as losses will never be entirely avoidable, choosing the right system ordering, $\topA$, is of utmost importance. 

\section{Conclusion}\label{conclusion}
In this work, we discussed a cascaded version of the all-optical coherent quantum noise cancellation setup proposed by Tsang and Caves \cite{Tsang2010,Wimmer2014}. Instead of introducing the anti-noise path directly into the optomechanical cavity, we considered an all-optical effective negative mass oscillator as a standalone system and removed the backaction noise of the positive mass oscillator by coupling both systems coherently via a strong drive field. Under the conditions \eqref{DCBSmatch}--\eqref{matchingsusc}, we then rediscovered the perfect cancellation of backaction noise. Afterwards, we discussed deviations from the ideal conditions, including losses and the influence of the system order. We saw that for mismatched measurement strengths, by choosing the cavity linewidth and coupling strength in a specific way, CQNC can be recovered in the high- or low-frequency regime, or even at a specific frequency $\omega_{*}\neq\omega_\m$. For losses and a relative mismatch of beam-splitter and down-conversion coupling, the system order will also affect the noise cancellation performance. Finally, we discussed the performance of our setup for a set of realistic parameters and showed that a quantum noise reduction of 20\% below the SQL is possible for the order $\topA$ in the low-frequency regime.

We thank Jonas Junker and Bernd Schulte for fruitful discussions regarding the experimental setup.
This research was funded by the Deutsche Forschungsgemeinschaft (Excellence Cluster QuantumFrontiers (EXC 2123 Project ID 390837967), SFB 1227 (DQ-mat, project A05), GRK 1991) and the Quantum- and Nano-Metrology (QUANOMET) initiative from VW-Vorab (ZN3294).

\begin{appendix}
\section{Details on the Hamiltonian \texorpdfstring{$H_\negm$}{Hnmo}}\label{appendixHnegm}
Here, we give some more details on the Hamiltonian $H_\negm$ in Eq.~\eqref{Hamnm}. The effective negative mass oscillator consists of two optical modes with different frequencies $\omega_\c$ and $\omega_\a$, coupled by a beam-splitter and down-conversion process. The whole Hamiltonian reads
\begin{align}
    H_\negm = H_0 + H_\mathrm{drive}+ H_\DC + H_\BS,
\end{align}
where the first term represents the free Hamiltonian of the optical modes,
\begin{align}
    H_0 = \omega_\c c^\dagger_\c c_\c + \omega_\a a^\dagger a.
\end{align}
The second term describes the laser drive, which drives the meter cavity mode $c_\c$, and is given by
\begin{align}
    H_\mathrm{drive} = i E (e^{-i\omega_\Li t} c^\dagger_\c - e^{i\omega_\Li t} c_\c).
\end{align}
Here, $\omega_\Li$ denotes the laser frequency and $E$ describes the laser field amplitude, which is given by $\lvert E \rvert = \sqrt{\kappa_\c P / \hbar \omega_\Li}$, with the laser power $P$ and cavity linewidth $\kappa_\c$.
The third term describes a two-mode squeezing process. In this, a pump field impinges on a non-linear crystal, and a pump photon of frequency $\omega_\mathrm{P}$ is converted into two photons of lower frequency. In the usual treatment of such processes, the pump is assumed to be a strong coherent field. In a rotating frame with respect to the pump frequency and linearized pump field \cite{Ou1992},  the Hamiltonian is then
\begin{align}
    H_\DC = g_\DC (a c_\c + a^\dagger c^\dagger_\c).
\end{align}
The coupling strength of the down-conversion process is given by $g_\DC= \Gamma l \frac{c}{L}$, where $l$ is the length of the crystal, $L$ is the cavity length, $c$ the speed of light and $\Gamma$ the gain parameter. We refer to \cite{Wimmer2014,Byer1977} for more details on the gain parameter.
The last term describes the beam-splitter interaction. It is given by
\begin{align}
    H_\BS = g_\BS (a^\dagger c_\c + c^\dagger_\c a),
\end{align}
where $g_\BS$ denotes the coupling strength of this process. For a generic beam-splitter, the strength is defined by $g_\BS=r c/L$ with $r$ the reflectivity of the beam-splitter. Alternatively, for a setup considered in Fig.~\ref{fig:setup}(a), where the two modes are not spatially separated, the beam-splitter interaction can be achieved with a wave plate \cite{Steinmeyer2019}. The strength is then given by
\begin{align}
    g_\BS = \frac{c}{L} \frac{\theta}{2}\sin 2 \tau,
\end{align}
with $\theta$ the wave plate angle and $\tau$ the delay.
Moving to a rotating frame with respect to the laser frequency $\omega_\Li$ and assuming a strong driving field, the Hamiltonian can be linearized, and we arrive at Eq.~\eqref{Hamnm}.

\section{Calculation of noise spectral densities}\label{appendixnoiseden}
We consider a general linear quantum system consisting of $n$ system variables, $k$ in- and outputs and $m$ bath variables. The input-output relations can be written as
\begin{align}\label{matrixinout}
    \xvec_\out = K_{\inp}^{\top} \xvec - \xvec_\inp
\end{align}
with a vector  $\xvec$ containing the $n$ system variables, and $\xvec_\inp$ and $\xvec_\out$ vectors for the $k$ inputs and outputs. The whole system is governed by the equations of motion
\begin{align}\label{matrixQLE}
    \Dot{\xvec}(t)= M_\sys \xvec(t) + K_\inp \xvec_\inp(t) + K_\bath \xvec_\bath(t),
\end{align}
where we have introduced the system matrix $M_\sys$ and input matrices $K_\inp$ and $K_\bath$ for the input and bath quadratures. The equations of motion \eqref{matrixQLE} can be solved in the Fourier domain, where $\Dot{\xvec}(t) = i \omega \xvec(\omega)$. It follows,
\begin{align}
\xvec = (i \omega \mathbb{1} - M_\sys)^{-1}\, (K_\inp \xvec_\inp+ K_\bath \xvec_\bath). \end{align}
Together with Eq.~\eqref{matrixinout}, we derive the output quadratures as
\begin{align}
\begin{split}
    \xvec_\out &= K_{\inp}^{\top} \xvec - \xvec_\inp\\
    &=  \left(K_{\inp}^{\top}(i \omega \mathbb{1} - M_\sys)^{-1} K_{\inp}-\mathbb{1}\right) \xvec_\inp\\ 
    & \quad +K_{\inp}^{\top}(i \omega \mathbb{1} - M_\sys)^{-1} K_{\bath} \xvec_\bath\\
    &= T_\inp \xvec_\inp + T_\bath \xvec_\bath\\
    &= \mathcal{T} \,\tilde{\xvec}_\inp,
\end{split}
\end{align}
where 
\begin{align}
    \tilde{\xvec}_\inp &= \bigl(\begin{smallmatrix}\xvec_\inp\\\xvec_\bath\end{smallmatrix}\bigr),\\
    \mathcal{T}&=\left(T_\inp, T_\bath\right).
\end{align}
From this, we can calculate the (symmetrized) spectral density matrix as
\begin{align}\label{specmatrix}
\begin{split}
    \delta(\omega-\omega')S_\out(\omega) &= \frac{1}{2}\langle \xvec_\out(\omega)\xvec^{\dagger}_\out(\omega')\rangle + \c.\c.\\
    &= \frac{1}{2}\langle \mathcal{T}(\omega)\tilde{\xvec}_\inp \tilde{\xvec}^{\dagger}_\inp\mathcal{T}^{\dagger}(-\omega')\rangle + \c.\c.\\
    &=\frac{1}{2}\langle \mathcal{T}(\omega) S_\inp \mathcal{T}^{\dagger}(-\omega')\rangle + \c.\c.,
\end{split}
\end{align}
with $S_\inp$ the input spectral density matrix.
Every sub-system in our setup has four system variables. Hence the system matrices $M_\sys$ and bath input matrices $K_\bath$ are all $4\times 4$-dimensional. The in- and output variables are the two quadratures of the laser light, making the input matrices $K_\inp$ $4 \times 2$-dimensional.

To model losses, the output quadratures are mixed with vacuum noise via a beam-splitter interaction,  which are then
\begin{align}\label{outputgeneral}
    \xvec_{\out}'= \underbrace{\begin{pmatrix}1 & 0 & 0 & 0\\
    0& 1 & 0& 0\end{pmatrix} \eta_{4\times4} \begin{pmatrix}\mathcal{T} & 0\\
    0 & \mathbb{1}\end{pmatrix}}_{\mathcal{T}_{\mathrm{\scriptscriptstyle{loss}}}}\begin{pmatrix}\tilde{\xvec}_\inp\\
   \xvec_\vac\end{pmatrix}.
\end{align}
The second matrix mixes the cavity output with vacuum, and the first matrix is the partial trace over the lost output port of the beam-splitter.

Finally, we need to cascade the two sub-systems. For this choose $\xvec_{\inp,2}=\xvec'_{\out,1}$, where the subscripts $1$ and $2$ stand for the first and second system. The total output quadratures are then given by
\begin{align}\label{totaltransfer}
\begin{split}
    \xvec'_{\out,2} &=\mathcal{T}^{\mathrm{\scriptscriptstyle{loss}}}_{2}\scalemath{0.85}{\begin{pmatrix} \mathcal{T}^{\mathrm{\scriptscriptstyle{loss}}}_1 & 0 & 0\\
    0 & \mathbb{1}_{4\times 4} & 0\\
    0 & 0 & \mathbb{1}_{2\times 2}
    \end{pmatrix}
    \begin{pmatrix} \xvec_\inp\\
    \xvec_{\bath,1}\\
    \xvec_{\vac,1}\\
    \xvec_{\bath,2}\\
    \xvec_{\vac,2}
    \end{pmatrix}}\\
    &=\mathcal{T}_{\mathrm{\scriptscriptstyle{total}}} \xvec_{\inp,\mathrm{\scriptscriptstyle{total}}}.
\end{split}
\end{align}

The equations of motion \eqref{QLEpm} for the optomechanical sensor imply the following matrices,
\begin{subequations}\label{matrixOMS}
\begin{align}
    M_\sys^\OMS &=\scalemath{0.75}{ \begin{pmatrix}-\frac{\kappa_{\om}}{2} & \Delta_\om & 0 & 0\\
    -\Delta_\om & -\frac{\kappa_{\om}}{2} & -g & 0 \\
    0 & 0 & 0 & \omega_\m \\
    -g & 0 & -\omega_\m & -\gamma_\m
    \end{pmatrix}},\\
    K^\OMS_{\inp} &= \scalemath{0.75}{\begin{pmatrix} \sqrt{\kappa^\inp_{\om}} & 0\\
    0 & \sqrt{\kappa^\inp_{\om}} \\
    0 & 0 \\
    0 & 0
    \end{pmatrix}},\\
    K^\OMS_{\bath}&=\scalemath{0.75}{\begin{pmatrix} \sqrt{\kappa^\bath_{\om}} & 0\\
    0 & \sqrt{\kappa^\bath_{\om}}\\
    0 & 0 \\
    0 & \sqrt{\gamma_{\m}}
    \end{pmatrix}}.
\end{align}
\end{subequations}
Similarly, the equations of motion  \eqref{QLEnm} for the  effective negative mass oscillator implies
\begin{subequations}\label{matrixNMO}
\begin{align}
    M_\sys^\NMO &= \scalemath{0.75}{\begin{pmatrix}-\frac{\kappa_{\c}}{2} & \Delta_\c & 0 &(g_\BS -g_\DC)\\
    -\Delta_\c & -\frac{\kappa_{\c}}{2} & -(g_\BS +g_\DC) & 0 \\
    0 & (g_\BS -g_\DC) & -\frac{\kappa_{\a}}{2} & \Delta_\a \\
   -(g_\BS +g_\DC) & 0 & -\Delta_\a & -\frac{\kappa_{\a}}{2}
    \end{pmatrix}},\\
    K^\NMO_{\inp} &= \scalemath{0.75}{\begin{pmatrix} \sqrt{\kappa^\inp_{\c}} & 0\\
    0 & \sqrt{\kappa^\inp_{\c}} \\
    0 & 0 \\
    0 & 0
    \end{pmatrix}},\\
    K^\OMS_{\bath}&=\scalemath{0.75}{\begin{pmatrix} \sqrt{\kappa^\bath_{\c}} & 0\\
    0 & \sqrt{\kappa^\bath_{\c}}\\
    \sqrt{\kappa_{\a}} & 0 \\
    0 & \sqrt{\kappa_{\a}}
    \end{pmatrix}}.
\end{align}
\end{subequations}

From these expressions, we calculate the total transfer matrix in Eq.~\eqref{totaltransfer} and together with the input spectral density,
\begin{align}
    S_\inp = \scalemath{0.85}{\begin{cases} \frac{1}{2}\,\mathrm{diag}\left(1,1,1,1,1,1,1,1,0,2 S_F \right)& \text{for }\topA \\
 \frac{1}{2}\,\mathrm{diag}\left(1,1,1,1,0,2 S_F,1,1,1,1 \right)& \text{for }\topB
  \end{cases}},
\end{align}
we obtain the output spectral density with Eq.~\eqref{specmatrix}. The spectral density of the added noise is then estimated from the phase component $S^{pp}_\out$, by dividing it by the coefficient of $S_F$.
\end{appendix}

\bibliography{cascCQNC}

\end{document}